\documentclass[twocolumn,pra,aps,showpacs]{revtex4}
\usepackage{bm}
\usepackage{mathrsfs}
\usepackage{amsmath}
\usepackage{amssymb}
\usepackage{graphicx}
\usepackage{amsfonts}
\usepackage{amsthm}
\usepackage{color}
\usepackage{dcolumn}
\usepackage{txfonts}

\begin{document}

\title{Preserving qubit coherence by dynamical decoupling}
\author{Wen YANG}
\address{Department of Physics, University of California San Diego, La Jolla, California 92093-0319, USA}
\author{Zhen-Yu WANG}
\author{Ren-Bao LIU}
\thanks{Email: rbliu@cuhk.edu.hk}
\affiliation{Department of Physics, The Chinese University of Hong Kong, Shatin, N. T., Hong Kong, China}

\begin{abstract}
In quantum information processing, it is vital to protect the
coherence of qubits in noisy environments.
Dynamical decoupling (DD), which applies
a sequence of flips on qubits and averages the
qubit-environment coupling to zero, is a promising strategy compatible
with other desired functionalities such as quantum gates.
Here we review the recent progresses in theories of dynamical decoupling and experimental
demonstrations. We give both semiclassical and quantum descriptions of the qubit decoherence
due to coupling to noisy environments. Based on the quantum picture, a geometrical
interpretation of DD is presented. The periodic Carr-Purcell-Meiboom-Gill DD and the
concatenated DD are reviewed, followed by a detailed exploration of the recently developed
Uhrig DD, which employs the least number of pulses in an unequally spaced sequence to suppress the
qubit-environment coupling to a given order of the evolution time. Some new developments and
perspectives are also discussed.
\end{abstract}


%%Quantum computation, 03.67.Lx
%%Decoherence:quantum mechanics, 03.65.Yz
%%Quantum entanglement, 03.65.Ud
%%Decoherence:quantum error correction, 03.67.Pp
%%Noise:fluctuation phenomena, 05.40.Ca
%%Pulse sequences, in NMR, 82.56.Jn

\maketitle

\section{Coherence and decoherence}

The power of quantum information
processing~\cite{Nielsen2000_Book}, the quantum
parallelism, comes from the superposition
principle of quantum mechanics. The building block of quantum technology,
a quantum bit (qubit), is a two-level system that can be
identified as a spin-1/2 with states $\left\vert \uparrow
\right\rangle $ and $\left\vert \downarrow \right\rangle $. The
ability of the qubit to be in a coherent superposition of
$\left\vert \uparrow \right\rangle $ and $\left\vert \downarrow
\right\rangle $,
\begin{align}
\left\vert \Psi \right\rangle =\cos \frac{\theta }{2}e^{-i\varphi
/2}\left\vert \uparrow \right\rangle +\sin \frac{\theta
}{2}e^{i\varphi /2}\left\vert \downarrow \right\rangle,
\end{align}%
enables the parallel processing of many pieces of classical information.
In order for this idea to work, the qubit has to faithfully maintain its quantum
state. Not only the populations $\cos ^{2}(\theta /2)$ and $\sin
^{2}(\theta /2)$ in the two states $\left\vert \uparrow
\right\rangle $ and $\left\vert \downarrow \right\rangle $, but also
the relative phase $e^{-i\varphi }$ between $\left\vert \uparrow
\right\rangle $ and $\left\vert \downarrow \right\rangle $ should be kept
at certain values.
Unavoidable couplings between the qubit and the environment
(hereafter referred to as bath) spoil the quantum state by
introducing uncontrolled evolution of the qubit. The populations and
phases are randomized and the qubit coherence is lost. This
decoherence problem is one of the most serious obstacles in
the roads towards scalable quantum information processing~\cite{DiVincenzo2000_FortschrPhys}.

The population randomization (i.e., relaxation) process involves energy
dissipation and therefore is subjected to the energy conservation condition. Thus
it can be suppressed by increasing the spin splitting of the qubit. In
contrast, the phase randomization (i.e., pure
dephasing) is a more serious issue, since this process does not involve energy
dissipation.

\subsection{Semiclassical picture of decoherence}

In the semiclassical picture, the pure dephasing of a qubit or a spin-1/2 is caused by the fluctuation of
a local classical field fixed at a given direction~\cite{Kubo1954_JPSJ,Anderson1954_JPSJ}.  The Hamiltonian of the qubit in the external field
including the random component is
\begin{align}
\hat{H}_{\mathrm{qubit}}=\left[\omega _{0}/2+ Z(t)\right]\hat{%
\sigma}_{z},
\end{align}%
where $\hat{\sigma}_{z}$ is the Pauli matrix for the qubit, $\omega_0$ is
the Zeeman splitting under the external field, and $2 Z(t)$ is the random field
resulting from the interaction with the bath. Let us consider a qubit initially
in a coherent superposition state
\begin{align}
\left\vert \psi (0)\right\rangle =C_{+}\left\vert \uparrow \right\rangle
+C_{-}\left\vert \downarrow \right\rangle,
\end{align}%
corresponding to a pure state density matrix%
\begin{align}
\hat{\rho}(0)=%
\begin{bmatrix}
\left\vert C_{+}\right\vert ^{2} & C_{+}C_{-}^{\ast } \\
C_{+}^{\ast }C_{-} & \left\vert C_{-}\right\vert ^{2}%
\end{bmatrix},
\end{align}%
in the basis $\left\vert \uparrow \right\rangle ,\left\vert \downarrow
\right\rangle $. At the end of the evolution, a random relative phase $%
\varphi (\tau )=2\int_{0}^{\tau }Z(t)dt$ between $\left\vert \uparrow
\right\rangle $ and $\left\vert \downarrow \right\rangle $ is accumulated in
the qubit wave function%
\begin{align}
\left\vert \psi (\tau )\right\rangle =C_{+}e^{-i\varphi (\tau )/2}\left\vert
\uparrow \right\rangle +C_{-}e^{i\varphi (\tau )/2}\left\vert \downarrow
\right\rangle,
\end{align}%
and the off-diagonal coherence of the resulting density matrix
\begin{align}
\hat{\rho}(\tau )=%
\begin{bmatrix}
\left\vert C_{+}\right\vert ^{2} & C_{+}C_{-}^{\ast }e^{-i\varphi (\tau )}
\\
C_{+}^{\ast }C_{-}e^{i\varphi (\tau )} & \left\vert C_{-}\right\vert ^{2}%
\end{bmatrix},
\end{align}%
becomes random. The ensemble average over all possible realizations
of the random noise $Z(t)$ gives the decay of the off-diagonal density matrix
elements, i.e., the decoherence of the qubit (or the depolarization of the spin-1/2 in the
plane perpendicular to the external field). The resulting qubit state is a
mixed state with vanishing off-diagonal coherence, since the noise-averaged
quantity $\left\langle e^{-i\varphi (\tau )}\right\rangle$
vanishes in the long time limit.

\subsection{Quantum theory of decoherence}

In the quantum picture~\cite{Yao2006_PRB}, the decoherence of a
qubit results from the qubit-bath entanglement,
which is established during the evolution of the interacting qubit-bath system.
The general pure dephasing Hamiltonian has the form
\begin{align}
\hat{H}_{\mathrm{dp}}=\hat{C}+\hat{\sigma}_{z}\otimes \hat{Z},
\label{PUREDP}
\end{align}%
where $\hat{C}$ is the interaction within the bath and $\hat{Z}$ is the bath operator representing
the quantum field on the qubit resulting from the qubit-bath interaction. Suppose the initial state
of the qubit-bath system has the form $\left\vert \Psi (0)\right\rangle =\left\vert \psi (0)\right\rangle \otimes
\left\vert J\right\rangle $ , i.e., a direct product
of the qubit state $\left\vert \psi (0)\right\rangle =C_{+}\left\vert
\uparrow \right\rangle +C_{-}\left\vert \downarrow \right\rangle $ and the
bath state $\left\vert J\right\rangle $. At the end of the evolution, an
entangled state is established as
\begin{align}
\left\vert \Psi (\tau )\right\rangle &=C_{+}\left\vert +\right\rangle
\otimes e^{-i(\hat{C}+\hat{Z})\tau }\left\vert J\right\rangle
+C_{-}\left\vert -\right\rangle \otimes e^{-i(\hat{C}-\hat{Z})\tau
}\left\vert J\right\rangle \nonumber \\
&\equiv C_{+}\left\vert +\right\rangle \otimes \left\vert
J^{(+)}(\tau )\right\rangle +C_{-}\left\vert -\right\rangle \otimes
\left\vert J^{(-)}(\tau )\right\rangle,
\end{align}%
and the off-diagonal coherence of the reduced density matrix of the qubit becomes bath-state-dependent
\begin{align}
\hat{\rho}(\tau )=%
\begin{bmatrix}
\left\vert C_{+}\right\vert ^{2} & C_{+}C_{-}^{\ast }\left\langle
J^{(-)}(\tau )|J^{(+)}(\tau )\right\rangle \\
C_{+}^{\ast }C_{-}\left\langle J^{(+)}(\tau )|J^{(-)}(\tau )\right\rangle &
\left\vert C_{-}\right\vert ^{2}%.
\end{bmatrix}.
\end{align}%
The off-diagonal qubit
coherence is reduced when the bath state overlap decreases
\begin{align}
L_{J}(\tau )\equiv \left\langle J^{(-)}(\tau )|J^{(+)}(\tau
)\right\rangle =\left\langle J|e^{i(\hat{C}-\hat{Z})\tau
}e^{-i(\hat{C}+\hat{Z})\tau }|J\right\rangle. \label{BathOverLap}
\end{align}
A transparent physical meaning of this formula is that the coherence
of the qubit decreases when the distinguishability  of the bath states
increases, or the quantumness of the qubit decays when it is gradually
``measured'' by the environment.

The decoherence in Eq.~(\ref{BathOverLap}) is caused by the quantum fluctuation
of the local field for a single bath state $|J\rangle$. At finite temperature,
the bath itself is in a thermal ensemble as $\sum_J P_J|J\rangle\langle J|$.
Ensemble average over the distribution of the initial
bath states $\left\vert J\right\rangle $ causes additional dephasing due to the
thermal fluctuation, referred to as inhomogeneous broadening in literature~\cite{Merkulov_2002}.

As an example, in a confined solid-state environment such as a quantum
dot, the most relevant source of decoherence at low temperature (a
few Kelvin) for an electron spin is the hyperfine interaction
with the lattice nuclear spins (which serve as the
bath)~\cite{Merkulov_2002,Fujisawa2002,Elzerman2004_Nature,Kroutvar2004_Nature}.
In a moderate ($\gtrsim 0.1$ Tesla in GaAs quantum dots) external
magnetic field, the electron spin relaxation is strongly
suppressed~\cite{Semenov2003,Deng2006,Shenvi2005a} and the coherence
decay is dominated by pure dephasing. Recently, a variety of quantum
many-body theories have been developed to evaluate the bath state
evolution $L_{J}(\tau )$ or its ensemble average, including the
density matrix cluster expansion~\cite{Witzel2006,Witzel2007},
the pair-correlation approximation~\cite{Yao2006_PRB},
the linked-cluster expansion~\cite{Saikin2007}, and the cluster
correlation expansion~\cite{Yang2008,Yang2009}. In the
pair-correlation approximation~\cite{Yao2006_PRB},
each pair-wise flip-flop of the nuclear spins is identified as an elementary excitation mode and is
taken as independent of each other. To study the higher order
correlations, the Feynman diagram linked-cluster expansion is
developed~\cite{Saikin2007}. The evaluation of higher-order linked-cluster expansion,
however, is tedious due to the increasing number and complexity of diagrams with
increasing the interaction order. The density matrix cluster expansion~\cite{Witzel2006,Witzel2007}
provides a simple solution  to include
the higher-order spin interaction effects beyond the
pair-correlation approximation (without the need to count or evaluate Feynman diagrams).
However, the accuracy problem (even
when the expansion converges) limits the cluster expansion to
applications in large spin baths. The cluster-correlation
expansion~\cite{Yang2008,Yang2009} bears the accuracy of the linked-cluster expansion (the
results are accurate whenever converge) and the simplicity of the
cluster expansion (without the need to count or evaluate Feynman
diagrams), while free from the large-bath restriction of the cluster
expansion.

\section{Suppressing decoherence by dynamical decoupling}

Since qubit decoherence results from uncontrolled evolution due to
the coupling between the qubit and the bath, a natural idea
to combat decoherence is to encode the qubit in a subspace immune to
noises from the environment (decoherence-free subspace~\cite{Duan:1997PRL,Lidar:1998PRL}),
which is made possible by symmetries of the interactions in certain physical systems.
Or alternatively, the coherence can be protected by dynamically eliminating the qubit-bath
coupling during the evolution (dynamical decoupling, referred to as DD for short).
The DD schemes were originated from the Hahn echo~\cite{Hahn1950_PR} and were developed
for high-precision magnetic resonance spectroscopy~\cite{Mehring_NMR,Rhim1970_PRL,haeberlen1976hrn}.
When the field of quantum computing was opened up, the idea of DD was introduced to protect qubit
coherence~\cite{Viola1998_PRA,Ban_DD,Zanardi:99PLA,Viola1999_PRL},
which stimulated numerous theoretical studies on extension and
applications of the DD approach to quantum computing~\cite{Viola2005_PRL,Kern2005_PRL,Khodjasteh2005_PRL,Khodjasteh:2007PRA,
Santos:2006PRL,Yao2007_RestoreCoherence,Liu2007_NJP,Witzel2007_PRBCDD,Zhang:2007PRB,Uhrig2007UDD,Cywinski2008}.
The recent experimental advances are also remarkable~\cite{Morton_NPhys06,Biercuk2009,Du2009}.

In the DD scheme, a sequence of pulses
is applied to flip the qubit and average the qubit-bath coupling to
zero during the evolution. It is a promising strategy due to its
compatiblility with other desired functionalities such as quantum
gates~\cite{Khodjasteh_DDgate,West_DDgate,Lidar_DDgate}.
The most general Hamiltonian describing the coupling between a qubit and a bath reads
\begin{align}
\hat{H}=\hat{C}+\hat{\sigma}_{x}\otimes \hat{X}+\hat{\sigma}_{y}\otimes \hat{%
Y}+\hat{\sigma}_{z}\otimes \hat{Z},  \label{HAMIL}
\end{align}%
where $\hat{\sigma}_{x/y/z}$ are the Pauli matrices for the
qubit, and $\hat{C},\hat{X},\hat{Y},$ and $\hat{Z}$ are bath operators. The
off-diagonal coupling $\left(\hat{\sigma}_{x}\otimes \hat{X}+\hat{\sigma}%
_{y}\otimes \hat{Y}\right)$ induces population relaxation. The diagonal
coupling $\hat{\sigma}_{z}\otimes \hat{Z}$ induces pure dephasing.

\subsection{Carr-Purcell-Meiboom-Gill DD}

For the sake of simplicity, we first consider the pure dephasing case
($\hat{X}=\hat{Y}=0$). In the absence of controlling pulses, the
evolution of the quantum state $\left\vert \Psi (\tau )\right\rangle
=\hat{U}_{0}\left\vert \Psi (0)\right\rangle $ of the coupled
qubit-bath system is driven by the free
propagator $\hat{U}_{0}\equiv e^{-i\hat{H}\tau }=e^{-i(\hat{C}+\hat{\sigma}%
_{z}\otimes \hat{Z})\tau }$.

The Hahn echo~\cite{Hahn1950_PR} is realized by a single instantaneous $\pi$ pulse applied at the
middle of the evolution to switch
the qubit states between $\left\vert \uparrow \right\rangle $ and $%
\left\vert \downarrow \right\rangle $,
\begin{align}
\left\vert \Psi (2\tau )\right\rangle =\hat{U}_{0}\hat{\sigma}_{x}\hat{U}%
_{0}\left\vert \Psi (0)\right\rangle,
\end{align}%
so that the propagator for the whole evolution from $0$ to $2\tau $ is $\hat{%
U}_{0}\hat{\sigma}_{x}\hat{U}_{0}=\hat{\sigma}_{x}\hat{U}_{1}$ with
\begin{align}
\hat{U}_{1} & \equiv \hat{\sigma}_{x}\hat{U}_{0}\hat{\sigma}_{x}\hat{U}%
_{0}=e^{-i(\hat{C}-\hat{\sigma}_{z}\otimes \hat{Z})\tau }e^{-i(\hat{C}+\hat{%
\sigma}_{z}\otimes \hat{Z})\tau }\nonumber
\\
& =e^{-i2\tau \lbrack \hat{C}+\hat{\sigma}%
_{z}\otimes \hat{Z}\cdot O(\hat{C}\tau )]}.
\end{align}
In the propagator, the qubit-bath coupling is eliminated in the first
order of the pulse interval $\tau$. By repeating the Hahn echo propagator
$\hat{U}_{1}$, the Carr-Purcell-Meiboom-Gill DD (CPMG)~\cite{Carr1954_CP,Meiboom1958}
can be constructed so as to preserve the coherence of the qubit for a long time.

The building block of CPMG consists of two instantaneous $\pi $ pulses
applied at $\tau $ and $3\tau $, respectively. At the end of the evolution $%
t=4\tau $, the state of the qubit-bath system is $\left\vert \Psi (4\tau
)\right\rangle =\hat{U}_{2}\left\vert \Psi (0)\right\rangle $, where the
propagator%
\begin{align}
\hat{U}_{2}& =\hat{U}_{0}\hat{\sigma}_{x}\hat{U}_{0}\hat{U}_{0}\hat{\sigma}_{x}%
\hat{U}_{0}=\hat{\sigma}_{x}\hat{U}_{1}\hat{\sigma}_{x}\hat{U}%
_{1} \nonumber \\
& =e^{-i4\tau \lbrack \hat{C}+\hat{\sigma}_{z}\otimes \hat{Z}\cdot O(\hat{C%
}^{2}\tau ^{2})]},
\label{CP}
\end{align}%
is obtained by embedding $\hat{U}_{1}$ into the basic structure $\hat{\sigma}%
_{x}\left( \cdots \right) \hat{\sigma}_{x}\left( \cdots \right) $.
The CPMG sequence of $2N$ pulses is obtained by
repeating the building block $\hat{U}_{2}$ for $N$ times. The propagator for
the whole evolution from $0$ to $T=4N\tau $ is%
\begin{align}
\hat{U}_{\mathrm{CPMG}}=\hat{U}_{2}^{N}=e^{-iT[\hat{C}+\hat{\sigma}%
_{z}\otimes \hat{Z}\cdot O(\hat{C}^{2}\tau ^{2})]}.
\end{align}
The qubit-bath coupling is eliminated up to the second order of the minimum
pulse interval $\tau $.

\subsection{Concatenated DD}

Note that in Eq.~(\ref{CP}), the building unit of CPMG can be viewed as a
nested application of the Hahn echo, which eliminates the qubit-bath coupling
to one order higher than the simple Hahn echo does. It was noticed in Ref.~\cite{Viola1999_PRL,Zanardi:99PLA}
that a mirror-symmetric arrangement of two DD sequences can decouple a quantum object to a higher order.
And furthermore, Ref.~\cite{Viola1999_PRL} mentioned the possibility of realizing DD to an arbitrary order by iterative construction.
Khodjasteh and Lidar proposed the first explicit concatenated DD (CDD) scheme~\cite{Khodjasteh2005_PRL,Khodjasteh:2007PRA}
to eliminate arbitrary qubit-bath coupling (including both diagonal and off-diagonal couplings) with an
intuitive geometrical understanding~\cite{Lidar_DD_geometry}. The idea of CDD
was further developed by incorporation of randomness into the sequence for improvement
of performance~\cite{Santos:2006PRL,Zhang:2007PRB}.
CDD schemes against pure dephasing were investigated for electron spin qubits in realistic solid-state
systems with nuclear spins as baths~\cite{Yao2007_RestoreCoherence,Witzel2007_PRBCDD}.
The advantage of CDD over the periodic DD sequences has been observed in experiments for
nuclear spin qubits in solid state environments~\cite{West_DDgate}.

The propagator for CDD is obtained by recursion
\begin{align}
\hat{U}_{n}=\hat{\sigma}_{x}\hat{U}_{n-1}\hat{\sigma}_{x}\hat{U}%
_{n-1}=e^{-i2^{n}\tau \lbrack \hat{C}+\hat{\sigma}_{z}\otimes \hat{Z}\cdot O(%
\hat{C}^{n}\tau ^{n})]},
\end{align}%
in which the qubit-bath coupling has been eliminated up to the $n$th order
of the minimum pulse interval $\tau $. By increasing the concatenation level
$n$, the qubit-bath coupling can be eliminated up to an arbitrary order of $\tau$.

For the most general qubit-bath Hamiltonian in Eq.~(\ref{HAMIL}),
the idea of concatenation can still be applied to eliminate both the
pure dephasing
term $\hat{\sigma}_{z}\otimes \hat{Z}$ and the relaxation term $\hat{%
\sigma}_{x}\otimes \hat{X}+\hat{\sigma}_{y}\otimes \hat{Y}$. In the absence
of controlling pulses, the evolution of the qubit-bath system is driven by
the free propagator $\hat{U}_{0}\equiv e^{-i\hat{H}\tau }$. The qubit-bath
coupling can be eliminated up to the first order of $\tau $ by the controlled
evolution~\cite{Khodjasteh2005_PRL}
\begin{align}
\hat{U}_{1} \equiv &\hat{U}_{0}\left[ \hat{\sigma}_{x}\hat{U}_{0}\hat{\sigma%
}_{x}\right] \left[ \hat{\sigma}_{y}\hat{U}_{0}\hat{\sigma}_{y}\right] \left[
\hat{\sigma}_{z}\hat{U}_{0}\hat{\sigma}_{z}\right] \nonumber \\
=&e^{-i\tau (\hat{C}+\hat{\sigma}_{x}\otimes \hat{X}+\hat{\sigma}%
_{y}\otimes \hat{Y}+\hat{\sigma}_{z}\otimes \hat{Z})}e^{-i\tau (\hat{C}+\hat{%
\sigma}_{x}\otimes \hat{X}-\hat{\sigma}_{y}\otimes \hat{Y}-\hat{\sigma}%
_{z}\otimes \hat{Z})} \nonumber \\
&\times e^{-i\tau (\hat{C}-\hat{\sigma}_{x}\otimes \hat{X}+\hat{\sigma}%
_{y}\otimes \hat{Y}-\hat{\sigma}_{z}\otimes \hat{Z})}e^{-i\tau (\hat{C}-\hat{%
\sigma}_{x}\otimes \hat{X}-\hat{\sigma}_{y}\otimes \hat{Y}+\hat{\sigma}%
_{z}\otimes \hat{Z})} \nonumber \\
=&e^{-i4\tau \lbrack \hat{C}+O(\alpha \beta \tau )]},
\end{align}%
where $\alpha $ denotes the norm of $\hat{C}$ and $\beta $ denotes the
norm of $\hat{X},\hat{Y},\hat{Z}$. Thus all the qubit-bath
coupling terms are eliminated in the first order of the minimum
pulse interval $\tau$. By concatenation, the propagator for the $n$th order CDD is%
\begin{align}
\hat{U}_{n} &\equiv \hat{U}_{n-1}\left[ \hat{\sigma}_{x}\hat{U}_{n-1}\hat{%
\sigma}_{x}\right] \left[ \hat{\sigma}_{y}\hat{U}_{n-1}\hat{\sigma}_{y}%
\right] \left[ \hat{\sigma}_{z}\hat{U}_{n-1}\hat{\sigma}_{z}\right] \nonumber \\
& =e^{-i2^{n}\tau \lbrack \hat{C}+O(\alpha \beta ^{n}\tau
^{n})+O(\alpha ^{2}\beta ^{n-1}\tau ^{n})+\cdots +O(\alpha ^{n}\beta
\tau ^{n})]},
\end{align}%
in which the qubit-bath coupling has been eliminated up to the $n$th order
of $\tau $. By increasing the concatenation level $n$, the qubit-bath
coupling can be eliminated up to an arbitrary order.

To eliminate the qubit-bath coupling to a given order $N$
of the evolution time, the number of instantaneous $\pi$
pulses scales exponentially with the order of CDD, namely, $N_{\mathrm{pulse}}=O(2^{N})$ for
eliminating the pure dephasing term and $N_{\mathrm{pulse}}=O(4^{N})$ for
eliminating all qubit-bath couplings. Since errors are inevitably introduced in each $\pi$ pulse,
it is desirable to minimize the number of controlling pulses used
to achieve a given order of decoupling.

\subsection{Uhrig DD}

Uhrig DD (UDD)~\cite{Uhrig2007UDD,Lee2008_PRL,UhrigNJP2008,Yang2008a}
is a remarkable advance in the DD theory.
UDD can eliminate the qubit-bath pure dephasing up to the $N$th order
of the evolution time using $N$ instantaneous $\pi$ pulses applied at
\begin{align}
T_{j}=T\sin ^{2}\frac{j\pi }{2(N+1)},\ \ \ \ \ (j=1,2,\ldots ,N),
\label{UhrigTj}
\end{align}%
during the evolution of the qubit-bath system from $0$ to $T$.
UDD is optimal in the sense that it uses the minimum number of control pulses
for a given order of decoupling.
Such pulse sequences for $N\le 5$ were first noticed by Dhar \textit{et al}
in designing control of the quantum Zeno effect~\cite{Dhar2006_PRL}.
The application of such sequences to DD was first proposed by Uhrig for a pure dephasing spin-boson
model~\cite{Uhrig2007UDD}. Then Lee, Witzel and Das Sarma
conjectured that UDD may work for a general pure dephasing model
with an analytical verification up to $N=9$~\cite{Lee2008_PRL}.
Later, computer-assisted algebra was used to verify the conjecture up
to $N=14$~\cite{UhrigNJP2008}. Finally, UDD was rigorously proved to be universal
for any order $N$~\cite{Yang2008a} and was also extended to the case of population relaxation~\cite{Yang2008a}. 
The performance bounds for UDD against pure dephasing were also established~\cite{Uhrigarxiv2010}.

By construction, UDD is optimal for a finite system (or a system with a hard cut-off in the spectrum)
in the \textquotedblleft high fidelity\textquotedblright\ regime where a short-time expansion of
the qubit-bath propagator converges. For the \textquotedblleft low
fidelity\textquotedblright\ regime, further theoretical
work~\cite{Pasini2010a,Cywinski2008} shows that UDD is optimal
when the noise spectrum has a hard cutoff, while CPMG performs better than CDD and UDD
when the noise has a soft cutoff or when the hard cutoff is not
reached by the spectrum filtering functions corresponding to the DD sequences.
The experimental investigations of UDD were carried out in
an array of $\sim 1000$ Be$^{+}$ ions~\cite{Biercuk2009}
and in irradiated malonic acid crystals~\cite{Du2009}.

\subsubsection{Spin-boson model: discovery of UDD}

The qubit-bath Hamiltonian $\hat{H}_{\mathrm{sb}}$ of the spin-boson pure
dephasing model~\cite{Uhrig2007UDD} corresponds to $\hat{C}=\sum_{i}\omega _{i}\hat{b}%
_{i}^{\dagger }\hat{b}_{i}$ and $\hat{Z}=\sum_{i}(\kappa_{i}/2)(\hat{b}%
_{i}^{\dagger }+\hat{b}_{i})$ in Eq.~(\ref{PUREDP}), where $\hat{b}_{i}$ is the bosonic
annihilation operator. For $N$
instantaneous $\pi $ pulses applied at $T_{1},T_{2},\cdots
,T_{N}\in \lbrack 0,T]$, the propagator for the evolution from
$0$ to $T$ is
\begin{align}
\hat{U}(T,0)=& \hat{U}_{0}(T-T_{N})\hat{\sigma}_{x}\hat{U}_{0}(T_{N}-T_{N-1})
\nonumber \\
& \cdots \hat{\sigma}_{x}\hat{U}_{0}(T_{2}-T_{1})\hat{\sigma}_{x}\hat{U}%
_{0}(T_{1}),
\end{align}%
where $\hat{U}_{0}(t)=e^{-i\hat{H}_{\mathrm{sb}}t}$ is the free propagator. $%
\hat{U}(T,0)$ can be written as $\hat{U}_{N}$ (for $N$ being even) or $\hat{%
\sigma}_{x}\hat{U}_{N}$ (for $N$ being odd) with
%\begin{widetext}
\begin{align}
\hat{U}_{N}  = & e^{-i(\hat{C}+(-1)^{N}\hat{\sigma}_{z}\otimes \hat{Z}%
)(T-T_{N})}e^{-i(\hat{C}+(-1)^{N-1}\hat{\sigma}_{z}\otimes \hat{Z}%
)(T_{N}-T_{N-1})}
\nonumber \\ &
\cdots e^{-i(\hat{C}-\hat{\sigma}_{z}\otimes \hat{Z}%
)(T_{2}-T_{1})}e^{-i(\hat{C}+\hat{\sigma}_{z}\otimes \hat{Z})T_{1}}
\nonumber \\
= & \frac{\hat{U}_{N}^{(+)}+\hat{U}_{N}^{(-)}}{2}+\hat{\sigma}_{z}\otimes
\frac{\hat{U}_{N}^{(+)}-\hat{U}_{N}^{(-)}}{2},
\end{align}%\end{widetext}
where $\hat{U}_{N}^{(\pm)}$ is obtained from $\hat{U}%
_{N}$ by replacing $\hat{\sigma}_{z}$ by $\pm 1$. In the $%
N$th order UDD, the positions $T_{1},T_{2},\cdots ,T_{N}$ of the $N$ pulses
are fixed by requiring that in the propagator $\hat{U}_{N}$, the qubit-bath
coupling should be eliminated up to the $N$th order, i.e.
\begin{align}
\delta \hat{U}(T)\equiv \hat{U}_{N}^{(+)}-\hat{U}_{N}^{(-)}=\hat{U}%
_{N}^{(-)}[(\hat{U}_{N}^{(-)})^{\dagger }\hat{U}^{(+)}-1]\sim O\left(T^{N+1}\right),
\end{align}%
or equivalently,
\begin{align}
\tilde{\delta}\hat{U}(T)\equiv (\hat{U}_{N}^{(-)})^{\dagger }\hat{U}^{(+)}-1 \sim O\left(T^{N+1}\right).
\end{align}
By exact diagonalization of the spin-boson
Hamiltonian, $\tilde{\delta}\hat{U}(T)$ has been evaluated as $\tilde{\delta}\hat{U}(T)=e^{2\hat{\Delta}(T)}$ with%
\begin{align}
\hat{\Delta}(T)=\sum_{j=0}^{N}(-1)^{j}\left[\hat{K}_{I}(T_{j})-\hat{K}_{I}(T_{j+1})\right],
\end{align}%
where we have defined $T_{0}\equiv 0$, $T_{N+1}\equiv T $, and
\begin{subequations}
\begin{align}
 \hat{K}_{I}(t) & \equiv e^{i\hat{C}t}\hat{K}e^{-i\hat{C}t}, \\
 \hat{K} & \equiv \sum_{i}\frac{\kappa_{i}}{2\omega_{i}}(\hat{b}_{i}^{\dagger }-\hat{b}_{i}).
\end{align}
\end{subequations}
The Taylor expansion
\begin{align}
\hat{K}_{I}(t)=\hat{K}+\sum_{p=1}^{\infty }\frac{(it)^{p}}{p!}\overset{p\text{%
\textrm{-fold commutator}}}{\overbrace{\left[ \hat{C},\cdots \left[ \hat{C},%
\left[ \hat{C},\hat{K}\right] \right] \right] }\equiv }\hat{K}%
+\sum_{p=1}^{\infty }\hat{K}_{p}t^{p},
\end{align}%
yields $\hat{\Delta}(T)=-\sum\limits_{p=1}^{\infty
}\hat{K}_{p}T^{p}\Lambda_{p}$, where
\begin{align}
\Lambda_{p} \equiv \sum\limits_{j=0}^{N}(-1)^{j}\left[\left(
\frac{T_{j+1}}{T}\right)^{p} - \left( \frac{T_{j}}{T}\right)^{p}
\right].
\label{lambdaP}
\end{align}%
Thus the condition $\tilde{\delta}\hat{U}(T)=O(T^{N+1})$ is equivalent to
$N$ coupled algebra equations%
\begin{align}
\Lambda_{p}=0,\ \ \ \ (p=1,2,\cdots ,N). \label{LambdaN}
\end{align}%
whose unique physical solution is the UDD sequence in
Eq.~(\ref{UhrigTj}). The UDD sequence is optimal in
that it uses the minimum number of pulses to make the first $N$ terms of
$\Lambda_{p}$'s vanish and eliminate the qubit-bath coupling up
to the $N$th order.

\subsubsection{Geometrical interpretation of decoherence and DD\label{sec:Appendix_GeoPic}}

Here we give a geometrical interpretation of decoherence and DD by
considering the spin-boson pure dephasing model, based on trajectories of
bath quantum states in the Hilbert space conditioned on the qubit states and
DD control. The pure dephasing qubit-bath Hamiltonian can be reformulated as
\begin{align}
\hat{H}\equiv\sum_{\pm}|\pm\rangle\langle\pm|\otimes \hat{H}_{\pm},
\end{align}
where
$\{|\pm\rangle\}$ denote the two eigenstates of the qubit, and the
bath operators $\hat{H}_{\pm}=\hat{C}\pm\hat{Z}$.
The qubit coherence is given by the overlap of bath states, as shown in
Eq.~(\ref{BathOverLap}).

The state of the bosonic bath can be described in the basis of coherent
states~\cite{Zhang:1990RevModPhys-CoherentStates}. The coherent state of the $l$th boson mode
is $|P_{l}\rangle\equiv
e^{P_{l}\hat{b}_{l}^{\dagger}-P_{l}^{*}\hat{b}_{l}}|0\rangle$ with
$P_{l}$ being a complex number. A coherent state $|P_{l}(t_{0})\rangle$
after a time of evolution under the Hamiltonians
$H_{l,\pm}=\omega_{l}\hat{b}_{l}^{\dagger}\hat{b}_{l}\pm\frac{1}{2}\kappa_{l}(\hat{b}_{l}^{\dagger}+\hat{b}_{l})$
is still a coherent state
\begin{align}
e^{-i\hat{H}_{l,\pm}(t-T_{0})}|P_{l}(T_{0})\rangle=|P_{l,\pm}(t)\rangle
e^{-i\theta_{\pm}},
\end{align}
where
\begin{align}
P_{l,\pm}(t)=[P_{l}(T_{0})-(\mp\frac{\kappa_{l}}{2\omega_{l}})]e^{-i\omega_{l}(t-T_{0})}\mp\frac{\kappa_{l}}{2\omega_{l}},
\end{align}
and the phase factor
$\theta_{\pm}(t)=\pm\int_{t_{0}}^{t}\Re[\frac{1}{2}\kappa_{l}P_{l}^{*}(t)]dt$.

\begin{figure}[t]
\includegraphics[width=0.9\columnwidth]{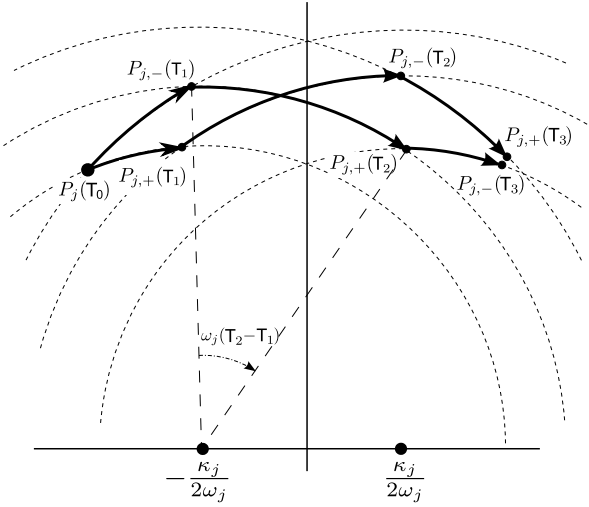}
\caption{\label{fig:The-bifurcated-trajectories}
The bifurcated trajectories
(solid curves) of $P_{j,\pm}(t)$ in the complex plane under qubit
flip control applied at $T_{1}=\frac{{T}}{4}$ and $T_{2}=\frac{3{T}}{4}$,
with the initial time $T_{0}=0$ and final time $T_{3}={T}$. The
arrows on the solid curves indicate the evolution directions.}
\end{figure}

As illustrated in Fig.~\ref{fig:The-bifurcated-trajectories}, 
the complex numbers $P_{l,\pm}(t)$, which represent the coherent states, are rotating clockwise about the points
$\mp\frac{\kappa_{l}}{2\omega_{l}}$ in the complex plane with an amplitude
of $|P_{l}(T_{0})|$ and an angular frequency $\omega_{l}$. The overlap of the bifurcated states
\begin{align} |\langle
P_{l,+}(t)|P_{l,-}(t)\rangle|=\exp(-|P_{l,+}(t)-P_{l,-}(t)|^{2}),
\label{eq:PathPOverlap}
\end{align}
is determined by their distance in the complex plane. We consider the case
that the initial bath state is a coherent state
$|J\rangle=\bigotimes_{l}|P_{l}(T_{0})\rangle$. Thus at time $t$ the
qubit coherence
\begin{align}
L_{J}(t)=\bigotimes_{l}|\langle P_{l,+}(t)|P_{l,-}(t)\rangle|,
\label{eq:PathCoherenceCoherState}
\end{align}
decreasing when the distance between $P_{l,+}(t)$ and $P_{l,-}$ in the complex plane is increased.
Since the bifurcated evolution of the bath is determined by the qubit
states $|\pm\rangle$, during the qubit-bath evolution, instantaneous
flips of the qubit states will cause the bath evolution pathways to exchange their rotation
centers. At some later time, the two bifurcated pathways could cross into each other,
upon which the qubit and the boson mode become disentangled. At this disentanglement 
point, the which-way information is erased, and therefore the qubit coherence is recovered.

Let the bifurcated bath states at time $T_{m-1}$ be denoted by
the complex numbers $\{P_{l,\mp}(T_{m-1})\}$. Suppose there is a qubit
flip applied at $t=T_{m-1}$. After an interval of evolution, the bath states will become
\begin{align}
P_{l,\pm}(T_{m})=[P_{l,\mp}(T_{m-1})\pm\frac{\kappa_{l}}{2\omega_{l}}]e^{-i\omega_{l}(T_{m}-T_{m-1})}\mp\frac{\kappa_{l}}{2\omega_{l}}.
\label{eq:PathPjRecursive}
\end{align}
We define the difference $\Delta_{m}^{l}\equiv P_{l,+}(T_{m})-P_{l,-}(T_{m})$.
By recursively using the initial condition
\begin{align}
P_{l,+}(T_{0})=P_{l,-}(T_{0})=P_{l}(T_{0}),
\end{align}
and Eq.~(\ref{eq:PathPjRecursive}), we have that after $N$
flips at times $T_{1}$, $T_{2}$, $\cdots$, $T_{N}$, the difference
\begin{align}
\Delta_{N+1}^{l}=i(-1)^{N+1}e^{-i\omega_{l}T_{N+1}}\kappa_{l}f(\omega_{l}),
\label{eq:PathDeltaNp1}
\end{align}
with
\begin{align}
f(\omega_{l})\equiv\frac{1}{i\omega_{l}}\sum_{j=0}^{N}(-1)^{j}(e^{i\omega_{l}T_{j+1}}-e^{i\omega_{l}T_{j}}).
\label{eq:Path-fw1}
\end{align}
Eqs.~(\ref{eq:PathPjRecursive}) and (\ref{eq:PathDeltaNp1}) give us
a geometrical interpretation of control of decoherence by qubit flips.
In Fig.~\ref{fig:The-bifurcated-trajectories}, we show the evolution
of $P_{j,\pm}(T_{m})$ for qubit flips occurring at $T_{1}=\frac{{T}}{4}$
and $T_{2}=\frac{3{T}}{4}$, with the total evolution time $T_{3}={T}$.

Note that the initial $P_{l}(T_{0})$ is canceled in the
expression of $\Delta_{N+1}^{l}$ in Eq.~(\ref{eq:PathDeltaNp1}). Thus
from Eqs.~(\ref{eq:PathPOverlap}) and (\ref{eq:PathCoherenceCoherState}),
the coherence $L_{J}$ is independent of the initial bath state
$|J\rangle=\bigotimes_{l}|P_{l}(T_{0})\rangle$.
By the expansion of $f(\omega_{l})$, we obtain
\begin{align}
\Delta_{N+1}^{l}=(-1)^{N+1}e^{-i\omega_{l}{T}}\frac{\kappa_{l}}{\omega_{l}}\sum_{n=1}^{\infty}\frac{(i\omega_{l}{T})^{n}}{n!}\Lambda_{n},
\end{align}
where $\Lambda_{n}$ is given by Eq.~(\ref{lambdaP}). The distance
$\Delta_{N+1}^{l}$ between $P_{l,\pm}(T_{N+1})$ is a small quantity
$\sim O\left({T}^{N+1}\right)$ if $\{\Lambda_{n}=0\}$ for $n\leq N$. Thus the
conditions for UDD are reproduced.

\subsubsection{Proof of universality of UDD against pure dephasing}

The proof of the universality (i.e., model independence)~\cite{Yang2008a}
of UDD is facilitated by the observation that to
eliminate the qubit-bath coupling to a given order, one needs only to eliminate the odd-power terms of the
coupling $\hat{\sigma}_{z}\otimes \hat{Z}$ in the perturbative expansion of the propagator,
since the even-power terms of $\hat{\sigma}_{z}\otimes \hat{Z}$ is a pure bath operator,
$(\hat{\sigma}_{z}\otimes \hat{Z})^{2m}=\hat{Z}^{2m}$, which does not
cause qubit decoherence. We will present the proof in the interaction picture
following Ref.~~\cite{Yang2008a}, which can be easily reformulated in other pictures~\cite{Uhrigarxiv2010}.

As discussed in the previous subsection, for the pure dephasing
Hamiltonian in Eq.~(\ref{PUREDP}) under the control of the $N$th order
UDD sequence, the propagator from $0$ to $T$ is given by
%\begin{widetext}
\begin{align}
\hat{U}_{N}=&e^{-i(\hat{C}+(-1)^{N}\hat{\sigma}_{z}\otimes \hat{Z}%
)(T-T_{N})}e^{-i(\hat{C}+(-1)^{N-1}\hat{\sigma}_{z}\otimes \hat{Z}%
)(T_{N}-T_{N-1})}
\nonumber \\ & \cdots e^{-i(\hat{C}-\hat{\sigma}_{z}\otimes \hat{Z}%
)(T_{2}-T_{1})}e^{-i(\hat{C}+\hat{\sigma}_{z}\otimes \hat{Z})T_{1}}.
\label{UN}
\end{align}
%\end{widetext}
Proof of the universality of UDD is equivalent to proving
\begin{align}
\hat{U}_{N}=\hat{U}_{N}^{(\mathrm{bath})}+O(T^{N+1}),
\end{align}%
where $\hat{U}_{N}^{(\mathrm{bath})}$ is a bath operator containing no qubit operators.
With the standard perturbation theory in the interaction picture, Eq.~(\ref{UN})
can be put in the time-ordered formal expression
\begin{align}
\hat{U}_{N}=e^{-i\hat{C}T}\hat{\mathscr
T}e^{-i\int_{0}^{T}F_{N}\left( t\right) \hat{\sigma}_{z}\otimes
\hat{Z}_{I}\left( t\right) dt},
\label{UpmTorder}
\end{align}%
where $\hat{\mathscr T}$ is the time-ordering operator, the modulation
function $F_{N}\left( t\right) \equiv \left( -1\right) ^{j}$ for $t\in \left[
T_{j},T_{j+1}\right] $ with $T_{0}\equiv 0$ and $T_{N+1}\equiv T$, and
\begin{align}
\hat{Z}_{I}(t)&\equiv e^{i\hat{C}t}\hat{Z}e^{-i\hat{C}t}=\sum_{p=0}^{\infty }%
\frac{(it)^{p}}{p!}\overset{p\text{-}\mathrm{folds\ commutator}}{\overbrace{%
\left[ \hat{C},\left[ \hat{C},\cdots \left[ \hat{C},\hat{Z}\right] \cdots %
\right] \right] }}\nonumber \\
& \equiv \sum_{p=0}^{\infty }\hat{Z}_{p}t^{p}.
\label{expansion}
\end{align}%
The propagator can be expanded into Taylor series%
\begin{align}
\hat{U}_{N}=e^{-i\hat{C}T}\sum_{n=0}^{\infty }(-i\hat{\sigma}%
_{z})^{n}\otimes \hat{\Delta}_{n}\equiv \hat{U}_{N}^{(\mathrm{even})}+\hat{U}%
_{N}^{(\mathrm{odd})},
\end{align}%
where%
%\begin{widetext}
\begin{align}
\hat{\Delta}_{n}  \equiv & \int_{0}^{T}F_{N}\left( t_{n}\right)
dt_{n}\int_{0}^{t_{n}}F_{N}(t_{n-1})dt_{n-1} \nonumber \\ &
\cdots
\int_{0}^{t_{2}}F_{N}(t_{1})dt_{1}\ \hat{Z}_{I}\left( t_{n}\right) \hat{Z}%
_{I}\left( t_{n-1}\right) \cdots \hat{Z}_{I}\left( t_{1}\right),
\label{DeltaN}
\end{align}%
%\end{widetext}
is a pure bath operator. Here
\begin{align}
\hat{U}_{N}^{(\mathrm{even})}=e^{-i\hat{C}T}\sum_{k=0}^{\infty }(-i)^{2k}%
\hat{\Delta}_{2k},
\end{align}%
consists of even powers of the qubit-bath coupling $\hat{\sigma}_{z}\otimes
\hat{Z}$ and therefore is a pure bath operator, which does not induce qubit
dephasing. The term consisting of the odd powers of the qubit bath coupling
\begin{align}
\hat{U}_{N}^{(\mathrm{odd})}=\hat{\sigma}_{z}\otimes e^{-i\hat{C}%
T}\sum_{k=0}^{\infty }(-i)^{2k+1}\hat{\Delta}_{2k+1},
\end{align}%
induces the qubit dephasing.
We just need to show $\hat{\Delta}_{2k+1}=O\left( T^{N+1}\right)$.

Using the expansion in Eq.~(\ref{expansion}), we have
\begin{align}
\hat{\Delta}_{n}=\sum_{\left\{ p_{j}\right\} }\left[
\hat{Z}_{p_{n}}\cdots
\hat{Z}_{p_{2}}\hat{Z}_{p_{1}}F_{p_{1},p_{2},\cdots
,p_{n}}T^{n+p_{1}+p_{2}\cdots +p_{n}}\right],
\end{align}%
where
\begin{align}
F_{p_{1},\cdots ,p_{n}}\equiv \int_{0}^{T}\frac{dt_{n}}{T}\cdots
\int_{0}^{t_{3}}\frac{dt_{2}}{T}\int_{0}^{t_{2}}\frac{dt_{1}}{T}%
\prod_{j=1}^{n}F_{N}\left( t_{j}\right) \left( \frac{t_{j}}{T}\right)
^{p_{j}},
\end{align}%
is a dimensionless constant independent of $T$. Now the problem is reduced
to proving
\begin{align}
F_{p_{1},p_{2},\cdots ,p_{n}}=0,  \label{PolyLemma}
\end{align}%
for $n$ being odd and $n+\sum_{j=1}^{n}p_{j}\leq N$. For this purpose, we
make the variable substitution $t_{j}=T\sin ^{2}(\theta _{j}/2)$ and define
the scaled modulation function
\begin{align}
f_{N}\left( \theta \right)  \equiv F_{N}\left( T\sin ^{2}(\theta /2)\right)
=\left( -1\right) ^{j},
\end{align}
for $\theta \in \left[ j\pi/(N+1),\left( j+1\right) \pi /(N+1)\right]$.
With
\begin{align}
\sin ^{2p}\frac{\theta }{2}\sin \theta
=(2i)^{-2p}\sum_{r=0}^{2p}C_{2p}^{r}\sin \left[ \left( p-r+1\right) \theta %
\right],
\end{align}%
we can write $F_{p_{1},p_{2},\cdots ,p_{n}}$ as a linear combination of
terms in the form
\begin{align}
f_{q_{1},\cdots ,q_{n}}\equiv \int_{0}^{\pi }d\theta _{n}\cdots
\int_{0}^{\theta _{3}}d\theta _{2}\int_{0}^{\theta _{2}}d\theta
_{1}\prod_{j=1}^{n}f_{N}\left( \theta _{j}\right) \sin \left(
q_{j}\theta _{j}\right),
\end{align}%
with $\left\vert q_{j}\right\vert \leq p_{j}+1$. Suffices it to show $%
f_{q_{1},q_{2},\cdots ,q_{n}}=0$ for odd $n$ and $\sum_{j=1}^{n}\left\vert
q_{j}\right\vert \leq N$. We notice that $f_{N}\left( \theta \right) $ is a periodic
function with a period of $2\pi /(N+1)$ and therefore can be expanded into Fourier series
\begin{align}
f_{N}\left( \theta \right) =\sum_{k=1,3,5,\cdots }\frac{4}{k\pi
}\sin \left[ k\left( N+1\right) \theta \right].  \label{fxFourier}
\end{align}%
The key feature of the Fourier expansion to be exploited is that it contains
only odd harmonics of $\sin [(N+1)\theta ]$. With the Fourier expansion, we
just need to show that
\begin{align}
\int_{0}^{\pi }d\theta _{n}\cdots \int_{0}^{\theta _{3}}d\theta
_{2}\int_{0}^{\theta _{2}}d\theta _{1}\prod_{j=1}^{n}\cos \left(
r_{j}\theta _{j}+q_{j}\theta _{j}\right) =0, \label{CosLemma0}
\end{align}%
for $n$ being odd, $r_{j}$ being an odd multiple of $(N+1)$, and $%
\sum_{j=1}^{n}\left\vert q_{j}\right\vert \leq N$. With the product-to-sum
trigonometric formula repeatedly used, it can be shown by induction that
after an even number of variables $\theta _{1},\theta _{2},\dots ,\theta
_{2k}$ have been integrated over, the resultant integrand as a function of $%
\theta _{2k+1}$ is the sum of cosine functions of the form
\begin{align}
\cos \left( R_{2k+1}\theta _{2k+1}+Q_{2k+1}\theta _{2k+1}\right),
\end{align}%
with $R_{2k+1}$ being an odd multiple of $(N+1)$ and $\left\vert
Q_{2k+1}\right\vert \leq \sum_{j=1}^{2k+1}\left\vert q_{j}\right\vert $. In
particular, the last step is
\begin{align}
\int_{0}^{\pi }\cos \left( R_{n}\theta _{n}+Q_{n}\theta _{n}\right)
d\theta _{n}.
\end{align}%
Since $R_{n}$ is an odd (non-zero, of course) multiple of $(N+1)$, and $%
\left\vert Q_{n}\right\vert \leq \sum_{j=1}^{n}\left\vert
q_{j}\right\vert \leq N $, we have $R_{n}+Q_{n}\neq 0$ and the
integral above must be zero. Thus Eq.~(\ref{PolyLemma}) holds. The
proof is done.

It should be noted that in the proof above, the perturbation-theoretical expansion requires that the
Hamiltonian of the bath have a finite norm, which means that the noise spectrum felt by the qubit has
a hard cutoff.

\subsubsection{Universality of UDD against population relaxation}
\label{sec_relax}

A straightforward corollary of Eq.~(\ref{PolyLemma}) is that UDD
can also be used to suppress population relaxation of the qubit.
Considering the most general qubit-bath Hamiltonian in
Eq.~(\ref{HAMIL}) and assuming that the UDD sequence consists of $N$
instantaneous $\pi $ pulses to rotate the
qubit around the $z$-axis, we aim to show that the relaxation of the qubit population in $%
\left\vert \uparrow \right\rangle $ and $\left\vert \downarrow \right\rangle
$ is eliminated up to $O( T^{N}) $. The propagator of the
qubit-bath evolution from $0$ to $T$ is%
%\begin{widetext}
\begin{align}
\hat{U}(T,0)=&\hat{U}_{0}(T-T_{N})\hat{\sigma}_{z}\hat{U}_{0}(T_{N}-T_{N-1})%
\nonumber \\ & \cdots \hat{\sigma}_{z}\hat{U}_{0}(T_{2}-T_{1})\hat{\sigma}_{z}\hat{U}%
_{0}(T_{1}),  \label{U_T_0}
\end{align}%
%\end{widetext}
where $\hat{U}_{0}(t)=e^{-i\hat{H}t}$ is the free propagator. $\hat{U}(T,0)$
can be written as $\hat{U}_{N}$ (for $N$ being odd) or $\hat{\sigma}_{z}\hat{%
U}_{N}$ (for $N$ being even) with
\begin{align}
\hat{U}_{N}=e^{-i[\hat{C}^{\prime }+(-1)^{N}\hat{D}](T-T_{N})}\cdots e^{-i(%
\hat{C}^{\prime }-\hat{D})(T_{2}-T_{1})}e^{-i(\hat{C}^{\prime }+\hat{D}%
)T_{1}}, \label{UN_RELAXATION}
\end{align}%
in which the Hamiltonian has been separated into $\hat{C}^{\prime }\equiv
\hat{C}+\hat{\sigma}_{z}\otimes \hat{Z}$ and $\hat{D}\equiv \hat{\sigma}%
_{x}\otimes \hat{X}+\hat{\sigma}_{y}\otimes \hat{Y}$. With the definition $%
\hat{D}_{I}\left( t\right) \equiv e^{i\hat{C}^{\prime }t}\hat{D}e^{-i\hat{C}%
^{\prime }t}$, the propagator can be formally expressed as
\begin{align}
\hat{U}_{N}=e^{-i\hat{C}^{\prime }T}\hat{\mathscr T}e^{-i\int_{0}^{T}F_{N}(t)%
\hat{D}_{I}(t)dt}=\hat{U}_{N}^{(\mathrm{even})}+\hat{U}_{N}^{(\mathrm{odd})},
\end{align}%
where%
\begin{align}
\hat{U}_{N}^{(\mathrm{even})}=e^{-i\hat{C}^{\prime }T}\sum_{k=0}^{\infty
}(-i)^{2k}\hat{\Delta}_{2k}^{\prime },
\end{align}%
consists of even powers of $\hat{D}$, and
\begin{align}
\hat{U}_{N}^{(\mathrm{odd})}=e^{-i\hat{C}^{\prime }T}\sum_{k=0}^{\infty
}(-i)^{2k+1}\hat{\Delta}_{2k+1}^{\prime },
\end{align}%
consists of odd powers of $\hat{D}$, with $\hat{\Delta}_{n}^{\prime }$
obtained from $\hat{\Delta}_{n}$ in Eq.~(\ref{DeltaN}) by replacing $\hat{Z}%
_{I}(t)$ by $\hat{D}_{I}(t)$. By expanding $\hat{D}_{I}(t)$ into
Taylor series [similar to Eq.~(\ref{expansion})]
\begin{align}
\hat{D}_{I}(t)=\sum_{p=0}^{\infty }\hat{D}_{p}t^{p},
\label{expansion2}
\end{align}%
the identity Eq.~(\ref{PolyLemma}) immediately gives $\hat{\Delta}%
_{2k+1}^{\prime }=O(T^{N+1})$. As a result, $\hat{U}_{N}^{(\mathrm{odd}%
)}=O(T^{N+1})$ and the propagator
\begin{align}
\hat{U}_{N}=\hat{U}_{N}^{(\mathrm{even})}+O(T^{N+1}),
\end{align}%
contains only even powers of $\hat{D}$ up to $O(T^{N})$. Since $\hat{D}$
contains only the Pauli matrices $\hat{\sigma}_{x}$ and $\hat{\sigma}_{y}$
and an even power of the two Pauli matrices $\hat{\sigma}_{x}^{n_{x}}\hat{%
\sigma}_{y}^{n_{y}}$ (with $n_{x}+n_{y}$ being even) is either unity or $i%
\hat{\sigma}_{z}$, the propagator
\begin{align}
\hat{U}_{N}=e^{-i\hat{H}_{\mathrm{eff}}(T)T+O(T^{N+1})},
\end{align}%
where the effective Hamiltonian $\hat{H}_{\mathrm{eff}}(T)=\hat{C}_{\mathrm{%
eff}}(T)+\hat{\sigma}_{z}\otimes \hat{Z}_{\mathrm{eff}}(T)$ contains only
pure dephasing term $\hat{\sigma}_{z}\otimes \hat{Z}_{\mathrm{eff}}(T)$
and commutes with $\hat{\sigma}_{z}$. Thus the $N$-pulse UDD eliminates the
population relaxation up to $O( T^{N}) $.

\subsubsection{Time-dependent Hamiltonians}

From the procedures following Eqs.~(\ref{expansion})
and (\ref{expansion2}), it is immediately observed that the proof above applies to time-dependent Hamiltonian
as long as a Taylor expansion of the Hamiltonian similar to those in Eqs.~(\ref{expansion})
and (\ref{expansion2}) exists (such as a Hamiltonian having analytical
time-dependence). Such a generalization was presented by Pasini and Uhrig~\cite{Pasini2010}.

\subsubsection{UDD with non-instantaneous pulses}

With the help of Eq.~(\ref{CosLemma0}), we realize that Eq.~(\ref{PolyLemma}%
) holds for more general modulation functions $F_{N}(t)$ as long as the
scaled modulation function $f_{N}(\theta )\equiv F_{N}\left( T\sin
^{2}(\theta /2)\right) $ contains only odd harmonics of $\sin [(N+1)\theta ]$
as in Eq.~(\ref{fxFourier}), i.e,
\begin{align}
f_{N}(\theta )=\sum_{k=0}^{\infty }A_{k}\sin \left[ (2k+1)(N+1)\theta \right],
  \label{odd_harmonics}
\end{align}%
with arbitrary coefficients $A_{k}$. Motivated by this observation, we try
to generalize UDD to the case of $\pi $ pulses with a finite duration.

For the case of UDD against general decoherence, we consider the control of
the qubit by an arbitrary time-dependent magnetic field $B(t)$
applied along a certain direction to protect the qubit coherence along this axis. Without loss of
generality, we take this direction as the $z$-axis. The general qubit-bath Hamiltonian
under DD control is
\begin{align}
\hat{H}(t)=\hat{C}+\hat{\sigma}_{x}\otimes
\hat{X}+\hat{\sigma}_{y}\otimes \hat{Y}+\hat{\sigma}_{z}\otimes
\hat{Z}+\frac{1}{2}\hat{\sigma}_{z}B(t).
\end{align}%
In the rotating reference frame following the qubit precession under the
magnetic field, the Hamiltonian becomes
\begin{align}
\hat{H}_{R}(t)=\hat{C}^{\prime }+\cos [\phi (t)]\hat{D}^{+}+\sin [\phi (t)]%
\hat{D}^{-},
\end{align}%
where the precession angle $\phi (t)=\int_{0}^{t}B\left( t^{\prime }\right)
dt^{\prime }$, $\hat{C}^{\prime }\equiv \hat{C}+\hat{\sigma}_{z}\otimes \hat{%
Z}$, $\hat{D}^{+}\equiv \hat{\sigma}_{x}\otimes \hat{X}+\hat{\sigma}%
_{y}\otimes \hat{Y}$, and $\hat{D}^{-}\equiv \hat{\sigma}_{x}\otimes \hat{Y}-%
\hat{\sigma}_{y}\otimes \hat{X}$. The propagator in the rotating reference
frame is
\begin{align}
\hat{U}_{N}=e^{-i\hat{C}^{\prime }T}\hat{\mathscr T}\exp \left(
-i\int_{0}^{T}\sum_{\lambda =\pm }F_{N}^{\lambda }(t)D_{I}^{\lambda
}(t)dt\right),
\end{align}%
with $F_{N}^{+}(t)=\cos [\phi (t)]$, $F_{N}^{-}(t)=\sin [\phi (t)]$, and $%
\hat{D}_{\mathrm{I}}^{\lambda }(t)=e^{i\hat{C}^{\prime }t}\hat{D}^{\lambda
}e^{-i\hat{C}^{\prime }t}$. Again we decompose $\hat{U}_{N}$ as the sum of $%
\hat{U}_{N}^{(\mathrm{even})}$ (which consists of even powers of $\hat{D}%
^{\pm }$) and $\hat{U}_{N}^{(\mathrm{odd})}$ (which consists of odd powers
of $\hat{D}^{\pm }$),
\begin{subequations}
\begin{align}
\hat{U}_{N}^{(\mathrm{even})} &=e^{-i\hat{C}^{\prime }T}\sum_{k=0}^{\infty
}\sum_{\lambda _{1}\cdots \lambda _{2k}}(-i)^{2k}\hat{\Delta}_{2k}^{(\lambda
_{1}\cdots \lambda _{2k})},  \\
\hat{U}_{N}^{(\mathrm{odd})} &=e^{-i\hat{C}^{\prime
}T}\sum_{k=0}^{\infty
}\sum_{\lambda _{1}\cdots \lambda _{2k+1}}(-i)^{2k+1}\hat{\Delta}%
_{2k+1}^{(\lambda _{1}\cdots \lambda _{2k+1})},
\end{align}
\end{subequations}%
where%
%\begin{widetext}
\begin{align}
\hat{\Delta}_{n}^{(\lambda _{1}\cdots \lambda _{n})}\equiv &
\int_{0}^{T}F_{N}^{\lambda _{n}}\left( t_{n}\right)
dt_{n}\int_{0}^{t_{n}}F_{N}^{\lambda _{n-1}}(t_{n-1})dt_{n-1}  \cdots
\nonumber \\ & \int_{0}^{t_{2}}F_{N}^{\lambda _{1}}(t_{1})dt_{1}\ \hat{D}_{I}^{\lambda
_{n}}\left( t_{n}\right) \hat{D}_{I}^{\lambda _{n-1}}\left( t_{n-1}\right)
\cdots \hat{D}_{I}^{\lambda _{1}}\left( t_{1}\right),
\end{align}%
%\end{widetext}
has a structure similar to $\hat{\Delta}_{n}$ in
Eq.~(\ref{DeltaN}). After expanding $\{\hat{D}_{I}^{\lambda
_{k}}(t_{k})\}$ into Taylor series,
the identity Eq.~(\ref{PolyLemma}) immediately gives $\hat{\Delta}%
_{2k+1}^{(\lambda _{1}\cdots \lambda _{2k+1})}=O(T^{N+1})$. Thus the
qubit decoherence along the $z$-axis is suppressed to $O\left( T^{N+1}\right) $] as long as the
scaled modulation function $f_{N}^{\pm }(\theta )\equiv F_{N}^{\pm }\left(
T\sin ^{2}(\theta /2)\right) $ contains only odd harmonics of $\sin
[(N+1)\theta ]$ as depicted in Eq.~(\ref{odd_harmonics}). This condition is
satisfied if and only if the scaled modulation functions $f_{N}^{\pm}(\theta )$ have the following symmetries:
\begin{enumerate}
\item periodic with period of $2\pi/(N+1)$;

\item anti-symmetric with respect to $\theta=j\pi/(N+1)$;

\item symmetric with respect to $\theta=(j+1/2)\pi/(N+1)$.
\end{enumerate}

\begin{figure}[t]
\includegraphics[width=0.9\columnwidth]{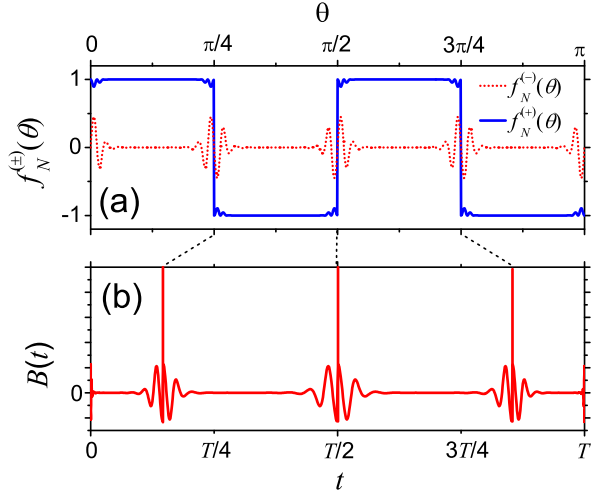}
\caption{(Extracted from Ref.~\cite{Yang2008a}, Copyright of the American Physical Society 2008)
An example of (a) the scaled modulation functions $%
f_{N}^{\pm }(\protect\theta )$ for the generalized 3rd order UDD control and
(b) the corresponding magnetic field $B(t)$. The dashed lines indicate the
correspondence between the sudden jumps of the modulation function $%
f_{N}^{+}(\protect\theta )$ in (a) and the sharp spikes as instantaneous $%
\protect\pi $-pulses in (b).}
\label{Pulses}
\end{figure}

The anti-symmetry condition requires $f_{N}^{\pm }(\theta )$ be either zero
or discontinuous at $\theta =j\pi /(N+1)$. But $f_{N}^{+}(\theta )$ and $%
f_{N}^{-}(\theta )$ cannot be simultaneously zero since they have to satisfy
the normalization condition
\begin{align}
\left[ f_{N}^{+}(\theta )\right] ^{2}+\left[ f_{N}^{-}(\theta
)\right] ^{2}=1,  \label{fxNorm}
\end{align}%
according to the definition of $F_{N}^{\pm }(t)$. So there must be sudden
jumps at least in one of two modulation functions at $\theta =j\pi /(N+1)$,
which means the controlling magnetic field $B(t)$ has to contain a $\delta $%
-pulse for $\pi $-rotation at $t=T_{j}$. With the initial conditions $%
f_{N}^{+}(0)=1$ and $f_{N}^{-}(0)=0$, one can choose the field such that $%
f_{N}^{-}(\theta )$ is continuous while $f_{N}^{+}(0)$ has sudden jumps
between $+1$ and $-1$ at $\theta =j\pi /(N+1)$. Thus, a generalized UDD
sequence can be chosen the following way: For $\theta \in \left[ 0,\pi
/(2N+2)\right] $, $f_{N}^{+}(\theta )$ can be arbitrary but sudden jumps
from $-1$ to $+1$ at $\theta =0$ and from $+1$ to $-1$ at $\pi /(2N+2)$, and
$f_{N}^{-}(\theta )$ is determined from the normalization condition as $%
f_{N}^{-}(\theta )=\pm \sqrt{1-\left[ f_{N}^{+}(\theta )\right] ^{2}}$. At
other regions, $f_{N}^{\pm }(\theta )$ are determined by the symmetry
requirements. The pulse amplitude $B(t)$ for the generalized UDD is
\begin{align}
B(t)=\frac{1}{F_{N}^{+}(t)}\frac{d}{dt}{F}_{N}^{-}(t)=\sum_{j=1}^{N}\pi
\delta \left( t-T_{j}\right) +B_{\mathrm{extra}}(t),
\end{align}%
which is a superposition of the instantaneous UDD pulses and an extra
component $B_{\mathrm{extra}}(t)$ being arbitrary but subject to the
symmetry requirements. The demand of $\delta $-pulses in the generalized UDD
is consistent with the previous finding in Ref.~\cite{PasiniPRA2008} that
the effect of an instantaneous $\pi $-pulse on the evolution of a qubit
coupled to a bath cannot be exactly reproduced by a pulse with a finite
magnitude. An example of the scaled modulation functions and the
corresponding magnetic field for the generalized 3rd order UDD control are
shown in Fig.~\ref{Pulses}. Notice that due to the variable transformation
from $\theta $ to $t $, the magnetic field $B(t)$ does not have the
symmetries as the scaled modulation functions $f_{N}^{\pm }(\theta )$. For
example, $B(t)$ is not periodic and the pulse at different time has
different width.

Obviously, the same argument holds for DD against pure dephasing
just by changing the rotation axis.

\subsubsection{UDD with pulses of finite amplitude}

In realistic experiments, the pulses have finite durations and amplitudes,
which introduces additional errors. There is a no-go
theorem which states that instantaneous $\pi$-pulses cannot be approximated by
pulses of finite amplitude and of short duration $\tau_{p}$ with
accuracy higher than the order $O(\tau_{p})$ without perturbing the
bath evolution~\cite{PasiniPRA2008,Pasini2008ShortPulse}. However,
as we have discussed above, the symmetric requirements of
$f_{N}(\theta)$ automatically guarantee the performance of UDD.
Uhrig and Pasini showed that by appropriately designing the pulses,
the qubit-bath Hamiltonian describing pure dephasing can be transformed into
the form~\cite{Uhrig2010}
\begin{align}
\hat{H}=\hat{C}+\tilde{F}_{N}(t)\hat{\sigma}_{z}\otimes\hat{Z}+O(\tau_{p}^{M}),
\end{align}
with the modulation function taking values from
$\{-1,0,1\}$. The scaled modulation function
$\tilde{f}_{N}(\theta)\equiv\tilde{F}_{N}\left(T\sin^{2}(\theta/2)\right)$
is designed to have the symmetries required in the previous proof
and therefore can be expanded by odd harmonics of $\sin[(N+1)\theta]$.
Thus, the decoherence is suppressed up to the order
$O(T^{N+1})+O(\tau_{p}^{M})$. This sequence can also suppress
longituddinal relaxation~\cite{Yang2008a,Uhrig2010}. An arbitrary
order $M$ of pulse shaping can be achieved by a recursive scheme
based on concatenation~\cite{Khodjasteh2010}.

\subsection{Comparison of decoupling efficiencies of UDD and CDD}

We consider a DD sequence of $N$ pulses, with a total evolution time $T$ and a minimum
pulse interval $\tau$. In CDD, the decoupling order is $n\sim \log_2 N$ and $\tau=T/2^{n}$. In UDD,
the decoupling order is $N$ and $\tau\sim T/N^2$. To be specific, our discussion is based on
the pure dephasing model. The situation for the general decoherence model is similar.
We compare the efficiencies of UDD and CDD in the two following scenarios:
\\
\\
{\bf Case I}:  The total evolution time $T$ is fixed.
The decoupling precision (defined as the effective
coupling under the DD control relative to the original one) in UDD was derived
as~\cite{Uhrigarxiv2010}
\begin{align}
\epsilon_{\rm UDD}\sim ||H||^N T^N/N!.
\end{align}
In CDD, it scales with the time and the decoupling order as~\cite{Khodjasteh2005_PRL,Liu2007_NJP,Lidar_DDgate}
\begin{align}
\epsilon_{\rm CDD}\sim \left(||H||T/\sqrt{N}\right)^{n}=\left(||H||T\right)^n/2^{n^2/2}.
\label{CDD_eff}
\end{align}
Thus with $T$ fixed, increasing the decoupling order and hence the number of pulses
always increases the decoupling precision. An arbitrarily high decoupling precision can be achieved simply
by choosing a sufficiently high order of DD (and correspondingly a sufficiently small pulse interval $\tau$).
In the high-fidelity regime ($T$ is small), the decoupling precision of UDD scales with the number of pulses
much faster than that of CDD. However, if we compare the efficiency of UDD and CDD of the same decoupling order $n$,
i.e., the $n$th order UDD (containing $n$ pulses) and the $n$th order CDD (containing $2^n$ pulses),
CDD has a much higher decoupling precision than UDD does ($T^n/2^{n^2/2}\ll T^n/n!$ for large $n$),
since the minimum pulse interval $\tau=T/2^n$ in CDD is much smaller than that in UDD ($\tau\sim T/n^2$).
For the same reason (namely, reduction of $\tau$), to achieve a given order of precision,
CDD indeed requires by far less than the seemingly exponential cost.
\\
\\
{\bf Case II}: The minimum pulse interval $\tau$ is fixed, which is a frequently encountered restriction in realistic experiments.
In this situation, increasing the order of DD leads to two competing effects~\cite{Uhrigarxiv2010,Viola_bounds}.
First, the qubit-bath coupling is eliminated to a higher order, which tends to increase the decoupling precision.
Second, the total evolution time $T$ increases and the bath has more time to inflict qubit decoherence.
Competition between these two effects leads to the existence of an optimal decoupling order,
beyond which further increasing the order of DD does not improve the decoupling precision any longer.
For a given minimum pulse interval $\tau$, the optimal order of UDD is~\cite{Uhrigarxiv2010}
\begin{align}
n_{\rm opt, UDD} \sim 1/\left(||H||)\tau\right),
\end{align}
and that of CDD is~\cite{Liu2007_NJP,West_DDgate,Lidar_DDgate}
\begin{align}
n_{\rm opt, CDD}\sim -\log_2\left(||H||\tau\right).
\end{align}
UDD has a much higher optimal level than CDD for a small minimum pulse interval, and therefore
the highest decoupling precision that can be achieved by UDD is much higher than that by CDD.

\subsection{Experimental progresses}

UDD was first realized in experiments by Biercuk
\textit{et al}. in an array of $\sim $1000 Be$^{+}$ ions in a
Penning ion trap~\cite{Biercuk2009,Biercuk2009a,Uys2009} with noises
mimicked by artificially introduced random modulation of the control fields.
The qubit states were realized using a ground-state electron-spin-flip
transition. Coherent qubit operations were achieved through a
quasi-optical microwave system. UDD was compared with CPMG in the ``low fidelity'' regime for various classical noise
spectra. The data show that UDD dramatically
outperforms CPMG for Ohmic noise [power spectrum $S(\omega )\propto\omega $]
with a sharp cutoff, while for the ambient magnetic field
fluctuations whose power spectrum $S(\omega )\propto 1/\omega ^{4}$
has a soft cutoff, UDD performs similarly to CPMG over the entire
range of accessible noise intensities, consistent with the theoretical
predictions~\cite{Pasini2010a,Cywinski2008}.

The first experimental realization of UDD against realistic noises
was achieved by Du \textit{et al} in a solid-state system, namely,
irradiated malonic acid single crystals. The spins of the radicals
in the crystals created by irradiation form an ensemble of
qubits. The nuclear spins, in samples with relatively low
concentrations of radicals, constitute the quantum bath which can be
considered as finite for the time-scales involved in the experiment and
therefore has a finite noise spectrum. The
pulsed electron paramagnetic resonance was used to demonstrate
the performance of UDD for preserving electron spin coherence
at temperatures from 50K to room temperature~\cite{Du2009}.
Using a seven-pulse UDD sequence, the
electron spin coherence time was prolonged from 0.04~ms to about
30~ms. The experimental data from different samples under various
conditions demonstrate that UDD performs better than CPMG in fighting against noises
from nuclear spins. The good agreement between the experiment and calculations based on
microscopic theories~\cite{Yang2008,Yang2009} enables the authors to identify the relevant
electron spin decoherence mechanisms as the electron-nuclear contact
hyperfine interaction and the electron-electron dipolar interaction.

\section{New developments}

\subsection{CUDD: Concatenation of UDD}

CDD can eliminate all the qubit-bath couplings
(including pure dephasing and population relaxation) up to an
arbitrary order $N$ at the cost of exponentially increasing number
(of the order $4^{N}$) of controlling pulses. In contrast, UDD
sequence uses the least number (i.e., $N$) of controlling pulses to
eliminate either pure dephasing or population relaxation (but not
both) to the desired order $N$. Based on a combination of CDD and
UDD, a new DD sequence (named CUDD) was proposed~\cite{Uhrig2009} to
suppress both the pure dephasing and the population relaxation to
order $N$ with a much less (of the order $N2^{N}$) number of pulses.
The essential idea of CUDD is to use the $N$th order UDD sequence
(instead of the free evolution) as the building block of CDD
sequence.

The propagator $\hat{U}_{N-\mathrm{UDD}}(T)$ for the qubit-bath
evolution driven by the general Hamiltonian Eq.~(\ref{HAMIL}) under
$N$th order UDD
sequence of $\pi $ rotation around the $z$ axis is%
\begin{align}
\hat{U}_{N-\mathrm{UDD}} &=e^{-i[\hat{C}^{\prime }+(-1)^{N}\hat{D}%
](T-T_{N})}\cdots e^{-i(\hat{C}^{\prime }-\hat{D})(T_{2}-T_{1})}e^{-i(\hat{C}%
^{\prime }+\hat{D})T_{1}} \nonumber \\
&=e^{-i\hat{H}_{\mathrm{eff}}(T)T+O(T^{N+1})},
\end{align}%
[see Eq.~(\ref{UN_RELAXATION})], where $\{T_{j}\}$ are given by Eq.~(\ref%
{UhrigTj}) and $\hat{H}_{\mathrm{eff}}(T)=\hat{C}_{\mathrm{eff}}(T)+\hat{%
\sigma}_{z}\otimes \hat{Z}_{\mathrm{eff}}(T)$ is a pure dephasing
Hamiltonian. The pure dephasing can be eliminated by embedding $\hat{U}_{N-%
\mathrm{UDD}}$ into the structure $\left[ \hat{\sigma}_{x}(\cdots )\hat{%
\sigma}_{x}\right] (\cdots )$. The propagator for the $m$th order
concatenation of $\hat{U}_{N-\mathrm{UDD}}$ is%
\begin{align}
\hat{U}_{N-\mathrm{UDD}}^{(m)}=\hat{\sigma}_{x}\hat{U}_{N-\mathrm{UDD}%
}^{(m-1)}\hat{\sigma}_{x}\hat{U}_{N-\mathrm{UDD}}^{(m-1)},
\end{align}%
with $\hat{U}_{N-\mathrm{UDD}}^{(0)}=\hat{U}_{N-\mathrm{UDD}}$. In the CUDD
scheme, $\hat{U}_{N-\mathrm{UDD}}^{(N)}$ eliminates both the pure dephasing
and population relaxation up to the $N$th order with $O(N2^{N})$ pulses.

\subsection{Near optimal DD by nesting UDD}

Recently West \textit{et al} proposed a near optimal
DD~\cite{West2010} obtained by nesting UDD sequences, dubbed quadratic DD (QDD),
to protect qubits against general noises. The inner
$N$th order UDD eliminates population relaxation and the outer
$N$th order UDD eliminates the pure dephasing, so that both pure
dephasing and population relaxation are eliminated up to the
$N$th order of the evolution time. Using $\hat{U}_{N-\mathrm{UDD}}^{(Z)}(\tau )$ to denote
the qubit-bath propagator driven by the general Hamiltonian
Eq.~(\ref{HAMIL}) under the $N$th order UDD sequence of $\pi $ rotation
around the $z$ axis, the
propagator of the $(N,M)$th order near optimal DD,%
%\begin{widetext}
\begin{align}
\hat{U}_{N}^{(M)}=&\hat{U}_{N-\mathrm{UDD}}^{(Z)}(T-T_{M})\hat{\sigma}_{x}%
\hat{U}_{N-\mathrm{UDD}}^{(Z)}(T_{M}-T_{M-1}) \nonumber \\ & \cdots \hat{\sigma}_{x}\hat{U}%
_{N-\mathrm{UDD}}^{(Z)}(T_{2}-T_{1})\hat{\sigma}_{x}\hat{U}_{N-\mathrm{UDD}%
}^{(Z)}(T_{1}),
\end{align}%
%\end{widetext}
is obtained from Eq.~(\ref{U_T_0}) by replacing the free propagator $\hat{U}%
_{0}(t)$ by $\hat{U}_{N-\mathrm{UDD}}^{(Z)}(t)$, where $\{T_{j}\}$
are given by Eq.~(\ref{UhrigTj}) with $N$ replaced with $M$. Thus
$\hat{U}_{N}^{(N)}$ eliminates both the pure dephasing and the
population relaxation up to the $N $th order using $O(N^{2})$
pulses. Numerical simulation shows that for a fixed number of
pulses, this DD sequence outperforms CDD and CUDD by exponential saving
of the number of the pulses and it is nearly
optimal for small $N$, differing from the optimal solutions by no
more than two pulses.

A proof of the QDD was attempted in
Ref.~\cite{Pasini2010} with the argument that after the inner
level of UDD control, the resulting effective Hamiltonian is time-dependent
and the outer level of UDD control applies to time-dependent Hamiltonians.
The effective Hamiltonian under the inner level of UDD control as defined in Ref.~\cite{Pasini2010},
however, is only piecewise analytical. It can be shown by some counter examples~\cite{WangZhenyu_Note}
that for a general piecewise analytical Hamiltonian taken as resulting from certain inner level of control,
it is not guaranteed that the outer level of decoupling can be realized to the desired order.
Thus it remains an open question to us why the nested UDD control works.

\subsection{Protecting multi-qubit states by UDD}

Mukhtar \textit{et al} recently showed~\cite{MultilevelUDD:2010} that by applying a sequence
of unitary operations
\begin{align}
\hat{P}_{\psi}=2|\psi\rangle\langle\psi|-I,
\end{align} on the
multi-level quantum system according to the timing of UDD, the
initial quantum state $|\psi\rangle$ is protected to the order of
$O\left(T^{N+1}\right)$. This operation was also given in Ref.~~\cite{Dhar2006_PRL}.

Obviously, we have $\hat{P}_{\psi}^{\dagger}=\hat{P}_{\psi}$. We
define the operators
\begin{subequations}
\begin{align}
\hat{C}& =(\hat{H}+\hat{P}_{\psi}\hat{H}\hat{P}_{\psi})/2, \\
\hat{Z} & =(\hat{H}-\hat{P}_{\psi}\hat{H}\hat{P}_{\psi})/2.
\end{align}
\end{subequations}
Then the system-bath Hamiltonian is separated into two parts
\begin{align}
\hat{H}=\hat{C}+\hat{Z},
\end{align}
where $\hat{C}$ commutes with
the operator $\hat{P}_{\psi}$ while $\hat{Z}$ anti-commutes with
$\hat{P}_{\psi}$, i. e,
\begin{subequations}
\begin{align}
\hat{P}_{\psi}\hat{C}\hat{P}_{\psi}& =\hat{C},\\
\hat{P}_{\psi}\hat{Z}\hat{P}_{\psi}& =-\hat{Z}.
\end{align}
\end{subequations}
By applying a sequence of $N$ operations $\hat{P}_{\psi}$ according
to the timing of UDD, the system-bath propagator reads
%\begin{widetext}
\begin{align}
\hat{U}_{N}=& \hat{P}_{\psi}^{N}e^{-i(\hat{C}+\hat{Z})(T-T_{N})}\hat{P}_{\psi}e^{-i(\hat{C}+\hat{Z})(T_{N}-T_{N-1})}
\nonumber \\ & \cdots\hat{P}_{\psi}e^{-i(\hat{C}+\hat{Z})(T_{2}-T_{1})}\hat{P}_{\psi}e^{-i(\hat{C}+\hat{Z})T_{1}}.
\end{align}
%\end{widetext}
Note that a final $\hat{P}_{\psi}$ pulse is required
for odd $N$. Similar to the procedure in the proof of the
universality of UDD, we rewrite the propagator as
\begin{align}
\hat{U}_{N}=e^{-i\hat{C}T}\hat{\mathscr T}
e^{-i\int_{0}^{T}F_{N}(t)\hat{Z}_{I}(t)dt},
\end{align}
where
$\hat{Z}_{I}(t)\equiv e^{i\hat{C}t}\hat{Z}e^{-i\hat{C}t}$
anti-commutes with $\hat{P}_{\psi}$. We separate $\hat{U}_{N}$ into
two parts
\begin{align}
\hat{U}_{N}\equiv\hat{U}_{N}^{(\mathrm{even})}+\hat{U}_{N}^{(\mathrm{odd})},
\end{align}
where
\begin{subequations}
\begin{align}
\hat{U}_{N}^{(\mathrm{even})}& =e^{-i\hat{C}T}\sum_{k=0}^{\infty}(-i)^{2k}\hat{\Delta}_{2k}, \\
\hat{U}_{N}^{(\mathrm{odd})}& =e^{-i\hat{C}T}\sum_{k=0}^{\infty}(-i)^{2k}\hat{\Delta}_{2k+1},
\end{align}
\end{subequations}
with
%\begin{widetext}
\begin{align}
\hat{\Delta}_{n}\equiv & \int_{0}^{T}F_{N}\left(t_{n}\right)dt_{n}\int_{0}^{t_{n}}F_{N}(t_{n-1})dt_{n-1}\cdots
\nonumber \\ &
\int_{0}^{t_{2}}F_{N}(t_{1})dt_{1}\
\hat{Z}_{I}\left(t_{n}\right)\hat{Z}_{I}\left(t_{n-1}\right)\cdots\hat{Z}_{I}\left(t_{1}\right).
\end{align}
%\end{widetext}
Obviously, $\hat{U}_{N}^{(\mathrm{even})}$ commutes with
$\hat{P}_{\psi}$, since it contains even powers of $\hat{Z}$.
Following the same arguments in the proof of UDD for qubit
dephasing, we conclude that
$\hat{U}_{N}^{(\mathrm{odd})}=O(T^{N+1}).$ Thus,
$\hat{P}_{\psi}\hat{U}_{N}=\hat{U}_{N}\hat{P}_{\psi}+O(T^{N+1})$,
which immediately indicate that the expectation value of
$\hat{P}_{\psi}$ and hence the quantum state $|\psi\rangle$ are
preserved up to $O(T^{N+1})$.

\section{Conclusions and perspectives}

In summary, we have given a review of recent progresses in protecting qubit coherence
by the dynamical decoupling schemes. The DD techniques are originated from the
magnetic resonance spectroscopy. The developments for quantum information technologies
can in turn advance the high-precision magnetic resonance spectroscopy. For example,
UDD has recently been applied in magnetic resonance imaging of tumors in animals~\cite{Warren_UDD}.
Extension of the spin coherence by DD may have important applications in nano-scale or even atomic scale
magnetometry~\cite{ZhaoNanNV}.

Remarkably, experiments have demonstrated the DD method as a particularly promising scheme
for protecting quantum coherence in quantum computing. As compared to the quantum error correction
schemes, the DD requires no auxiliary qubits and can be integrated naturally with the quantum gates
without extra hardware overhead. However, the DD approach has a shortcoming in that it
works only for slow baths or for non-Markovian noises, in the sense that the characteristic separation
time of the DD sequence is required to be shorter than or at least comparable to the inverse of
the characteristic width of the noise spectrum. The quantum error correction scheme has no
such requirements. In dealing with errors in quantum computing due to spontaneous emission,
combination of DD and quantum error correction was proposed~\cite{Lidar_DD_QEC}. It is conceivable that in future
quantum computing, the non-Markovian noises be decoupled by DD and the remaining Markovian noises be coped with
by quantum error correction.
In general, for a multi-qubit system coupled to both Markovian and non-Markovian noises,
a combination of the two paradigmatic error-countering methods provides a complete picture for scalable
quantum computing~\cite{Lidar_DDgate}.

In the present research of DD, mostly the pulses are assumed instantaneous with only a few exceptions.
Two important issues are under intensive research, and some remarkable results
have emerged recently~\cite{Khodjasteh:2007PRA,Khodjasteh2010,Khodjasteh_DDgate,Lidar_DDgate,Uhrigarxiv2010}.
One is how to extend the DD to implement high-fidelity quantum gates or hybrid DD with
quantum gates. Can some ideas be borrowed from DD for realizing dynamical control resilient to noises?
Such an issue was previously addressed in simulation of quantum processors with DD approaches~\cite{Wojcan02PRA}.
Recently, encouraging progresses have been made toward hybridization of quantum gates and
DD~\cite{Khodjasteh_DDgate,West_DDgate,Lidar_DDgate}. Another issue is how to design a quantum gate
(such as a qubit flip, which is required in DD) optimally in the presence of environmental noises.
Various optimization schemes have been invented for suppressing/minimizing the noise effect to a certain
order~\cite{Gordon_PRL08,Wilhem_PRL09,Khodjasteh09PRL,Clausen_PRL10}. Ref.~\cite{Khodjasteh2010} has established
a systematic method to achieve an arbitrary order of precision based on iterative construction of
finite-amplitude pulses. It is of interest to ask whether and how the pulse shaping for quantum
gates with an arbitrary order of precision can be systematically constructed without iteration,
with the development from CDD to UDD being an inspiring example.

% ----------------------------------------------------------------
%\bibliographystyle{amsplain}
%\bibliography{}
\begin{acknowledgements}
The work was supported by Hong Kong RGC/GRF CUHK402209.
We are grateful to D. Lidar for critical reading and useful comments.
\end{acknowledgements}

%\bibliography{references}

\end{document}